\documentclass[%
 reprint,
 amsmath,amssymb,
 aps,
]{revtex4-1}

\usepackage{graphicx}% Include figure files
\usepackage{dcolumn}% Align table columns on decimal point
\usepackage{bm}% bold math
\usepackage{lineno}
\begin{document}

%\preprint{APS/123-QED}

\title{ATLAS Tile Calorimeter Readout Electronics Upgrade Program for the High Luminosity LHC}%

\author{A. S. Cerqueira on Behalf of the ATLAS Tile Calorimeter Group}
\affiliation{Electrical Engineering Department, Federal University of Juiz de Fora}
%\affiliation{Authors' institution and/or address}
\collaboration{ATLAS Collaboration}

%\affiliation{Second institution and/or address}
%\affiliation{Third institution}

\date{\today}% It is always \today, today,


%\linenumbers
\begin{abstract}
The Tile Calorimeter (TileCal) is the hadronic calorimeter covering the most central region of the ATLAS experiment at LHC. The TileCal readout consists of about 10000 channels. The ATLAS upgrade program is divided in three phases: The Phase~0 occurs during 2013-2014, Phase~1 during 2018-1019 and finally Phase~2, which is foreseen for 2022-2023, whereafter the peak luminosity will reach 5-7 x 10$^{34}$ cm$^2$s$^{-1}$ (HL-LHC). The main TileCal upgrade is focused on the Phase~2 period. The upgrade aims at replacing the majority of the on- and off-detector electronics so that all calorimeter signals are directly digitized and sent to the off-detector electronics in the counting room. All new electronics must be able to cope with the increased radiation levels. An ambitious upgrade development program is pursued to study different electronics options. Three options are presently being investigated for the front-end electronic upgrade. The first option is an improved version of the present system built using commercial components, the second alternative is based on the development of a dedicated ASIC (Application Specific Integrated Circuit) and the third is the development of a new version of the “QIE” (Charge Integrator and Encoder)  based on the one developed for Fermilab. All three options will use the same readout and control system using high speed (up to 40~Gb/s) links for communication and clock synchronization. For the off-detector electronics a new back-end architecture is being developed. A demonstrator prototype read-out for a slice of the calorimeter with most of the new electronics, but still compatible with the present system, is planned to be inserted in ATLAS already in mid 2014 (at the end of the Phase~0 upgrade).
\end{abstract}

\maketitle

%\tableofcontents

\section{Introduction}

\begin{figure}
\begin{center}
\includegraphics[width=5cm]{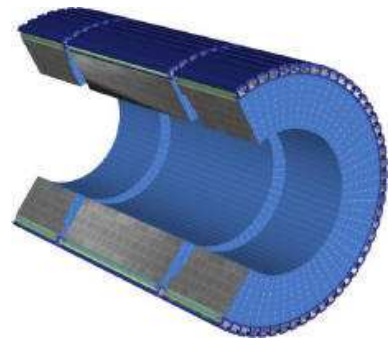}
\end{center}
\caption{\label{tilecal}. Illustration of the Tile Calorimeter.}
\end{figure}

TileCal is a sampling calorimeter composed of steel plates (tile shape) as absorber material interleaved with plastic scintillating plates as sampling material. It is divided in a central barrel (covering $|\eta|<1.0$) and two extended barrels (covering $0.8<|\eta|<1.7$), where each part is formed by 64 modules in order to complete the entire cylinder (see FIG. \ref{tilecal}).

When high energy particles interact with the steel, showers of lower energy particles are created, which in turn produce
light when passing through the scintillating tiles. The light is transmitted through wavelength shifting fibers to
photomultipliers (PMTs), which convert the light into electrical signals. Adjacent tiles and their WLS fibers are
grouped together to form TileCal cells. For each cell the fibers are read-out from two sides by two PMTs. Thus, each cell is read out via two different electrical signal paths. The central barrel modules are divided in up to 48
cells each, while the extended barrels modules are divided in 14 cells. Therefore, TileCal is comprised of more than 10000 readout
channels.

The TileCal front-end electronics, located inside the outermost part of the modules, is responsible for processing the
PMT signals and transmitting them to the back-end electronics, which is responsible for calorimeter signal
acquisition.

Currently, ATLAS is investigating necessary changes in order to prepare for the proposed high luminosity upgrade of the
LHC during the next decade (HL-LHC). Several detector components should be replaced (e.g. front-end electronics of the
calorimeters) and major changes on the trigger system are required to cope with the new luminosity
requirements.

In 2013-2014, the first long shutdown (LS1) at LHC is planned and will allow LHC to continue its operation reaching nominal peak
luminosity of $10^{34}$~cm$^2$s$^{-1}$ (Phase~0). A second long shutdown (LS2) is planned for 2018-2019 to prepare for
peak luminosities around $2-3\times10^{34}$~cm$^2$s$^{-1}$ corresponding to 55 to 80 interactions per bunch-crossing
with 25~ns bunch interval, well beyond the initial design goals (Phase~1). Finally, the Phase~2 upgrade is foreseen for
2022-2023  after which the peak luminosity will reach $5-7\times10^{34}$~cm$^2$s$^{-1}$ (HL-LHC). With luminosity leveling, the average luminosity will increase by a factor of 10.

This paper aims to present the TileCal front-end and back-end electronics upgrade program for the HL-LHC.
Next section
 presents the TileCal electronics and its upgrade plans for the HL-LHC. Section \ref{upgradea} presents details about the front-end and back-end upgrade projects. Finally, the
conclusions are stated in section \ref{conclusions}.

\begin{figure*}[t]
\begin{center}
\includegraphics[width=17cm]{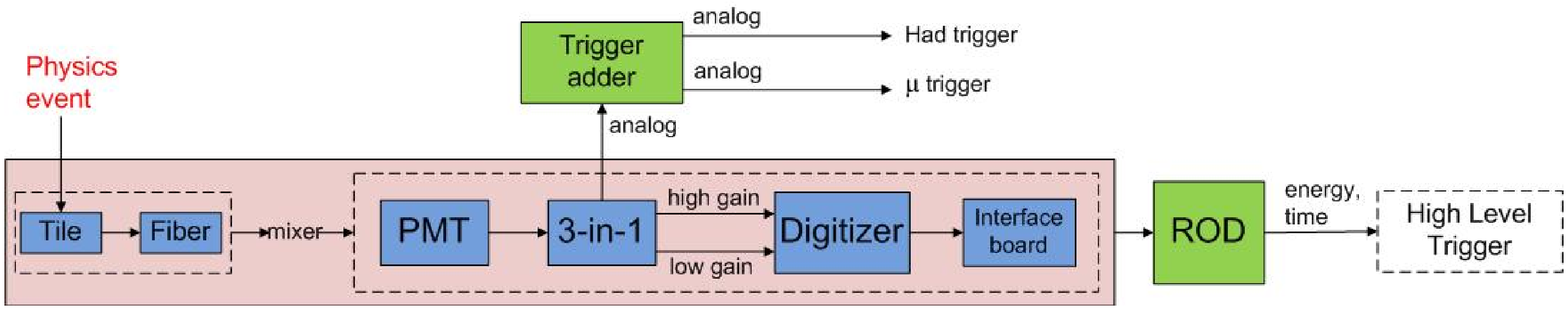}
\end{center}
\caption{\label{readoutchain}TileCal Signal Chain.}
\end{figure*}

\begin{figure*}[t]
\begin{center} 
\includegraphics[width=17cm]{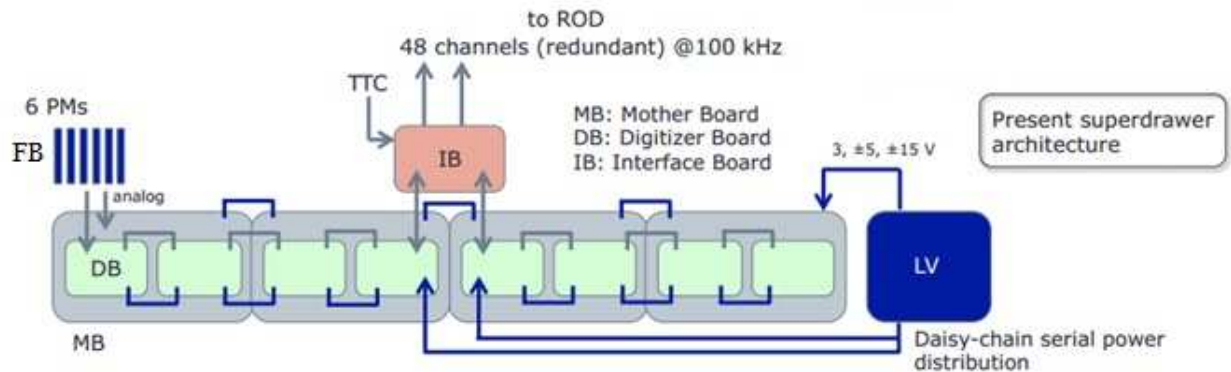} 
\end{center} 
\caption{\label{arch}Current archtecture of the TileCal electronics drawer.} 
\end{figure*}

\section{TileCal Electronics and Upgrade Plans for High Luminosity}
\label{upgradep} % More Details on Tilecal and the upgrade plans 

\subsection{TileCal Electronics}

The current TileCal signal chain can be seen in FIG.~\ref{readoutchain}. The light produced by energy depositions in the
detector is collected by optical fibers and sent to the light mixers, where several fibers are grouped together in order
to form the detector cells. The light is converted to an electrical signal in the photomultiplier tube (PMT) and is processed
by the 3-in-1 card (Front-End Board), which is responsible for signal conditioning and amplification providing three
analog signals as outputs, two for the detector readout (high and low gain) and another for triggering purpose. The low
and high gain signals are then digitized at 40~MHz by 10 bit Analog to Digital Converters (ADCs) in the Digitizer
Boards. Digital signals of all calorimeter cells in a module are merged and formatted into packages and sent via high speed optical links (Interface Board) that connect the on- and off-detector electronics. The
on-detector electronics is located in the outermost part of the TileCal module, in the electronics ``drawers''.

In the back-end electronics, the main component is the Read-Out Driver (ROD) which performs
preprocessing and gathers data coming from the front-end electronics at a maximum level 1 trigger rate of 100~kHz.
The ROD sends these data to the Read-Out Buffers (ROB) in the second level trigger \cite{trigger}.

FIG.~\ref{readoutchain} also shows the signal path to the first level trigger. The TileCal first level trigger signal
is produced by analog summation of up to six signals on the Trigger Board \cite{adder} and its analog output is sent to
the level one receiver by means of long twisted pairs cables (around 70~m). The first level trigger reduces the event rate from 40~MHz
to a maximum of 100~kHz.

The current archtecture of the TileCal electronics drawer can be seen in FIG.~\ref{arch}. It is possible to see the low voltage distribution, the four mother boards sections (MB), the eight digitizer boards (DB) along the drawer and the single interface board (IB) with two optical links for redundance, sending data to the back-end electronics at a maximum of 100~kHz rate.

The TileCal readout electronics must be modified to cope with the HL-LHC requirements. The luminosity increase results in higher radiation levels and higher data rates for the readout electronics since the events are larger, but the aim is to keep the first level trigger rate around the same maximum level, i.e. 100~kHz. To achieve this goal, the trigger system requires more information and more processing power to implement more complex algorithms. The new first level trigger will require full granularity digital information from the ATLAS calorimeters, differing from the analog calorimeter signals with reduced granularity in the current system.

\subsection{TileCal Upgrade Plans for the HL-LHC}

The main goals of the TileCal electronics upgrade program for high luminosity are the replacement of the aging electronics (2022-2023), the increase of radiation tolerance, the improvement of system reliability (less connectors - split Main Board design mitigates against single point failure causing loss of cell), to increase data precision and to improve the level one trigger system by the availability of the full detector resolution and improved Signal to Noise Ratio (SNR).

A new on-detector electronics architecture is under design and is being incrementally tested, where three different
design options for the new TileCal front-end board are under evaluation. The increased event rate also requires larger currents in the PMT voltage divider chains.
New active dividers and a new high voltage power supply are under development.
Concerning the off-detector electronics, a ``super'' Read-Out Driver (sROD) is being designed.
%The details about these projects will be described later in Section \ref{upgradea}.

Along with the development of new electronics a modification of the TileCal mechanics is being considered. The initial
development work has been carried out by Laboratoire de Physique Corpusculaire in Clermont-Ferrand (LPC). The aim is to
split the present drawers into two ``mini-drawers''. This is compatible with the new electronics architecture. The
mini-drawers will simplify handling of the drawers and improve the access to the TileCal electronics since it will be
easier to open the detector to replace mini-drawers. Practical solutions to insertion and cooling and electrical
connections are being tested with different prototypes.

The TileCal electronics upgrade program is making progress where an early prototype of the new system had already
been tested with one version of the new front-end board. The other two front-end options are being designed and are
scheduled for radiation test this year. In the end of 2013, one upgraded drawer should be ready for tests together with
the sROD. This new ``Demonstrator'' drawer will be backward compatible with the present system so that it can be seamlessly
installed in the present system. At the end of the first long shutdown, one Demonstrator drawer should be installed in the
detector in order to be tested under production conditions. If it proves to operate well three additional Demonstrator
drawers will be installed in later Christmas shut downs. Between 2014-2018 extensive tests should also be performed with
the three versions of the new system in test beam after which the decision about the front-end board to choose should be taken.
The production of new electronics will take place between 2019-2020. Finally, the installation is foreseen
for 2022-2023.

\section{On-Dectector and Off-Detector Electronics Upgrade Details}
\label{upgradea}

\subsection{On-Detector Electronics}

\begin{figure*}
\begin{center}
\includegraphics[width=16cm]{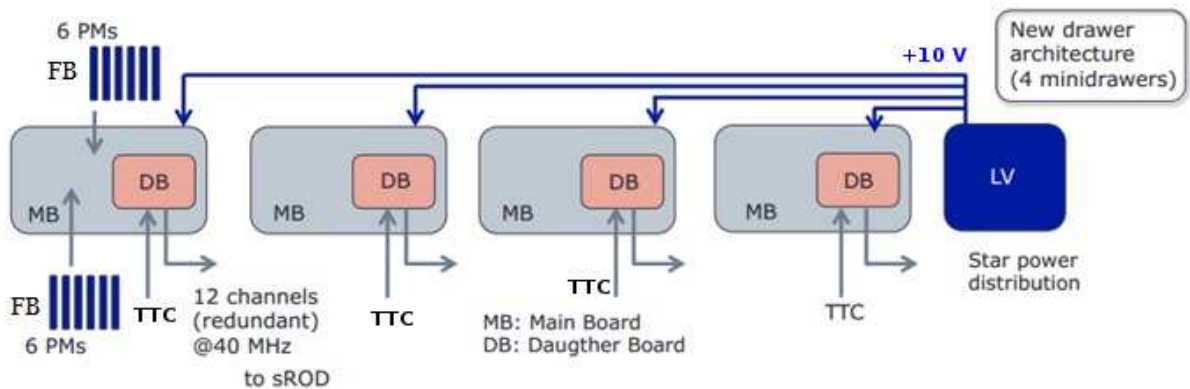}
\end{center}
\caption{\label{newarch}New drawer archtecture.}
\end{figure*}

The foreseen archtecture for the new on-detector electronics can be seen in FIG.~\ref{newarch}. The new readout
electronics is composed of: new Front-End boards (FE) that provide conditioning for the PMT signals as well
amplification, digitization (depending on the Front-End option) and calibration functionalities; new Main Boards (MB)
providing digitization (depending on the Front-End board) and control; and new Daughter Boards (DB) which provide data
processing and interface with the back-end electronics via optical links at up to 40~GHz rate with redundance. The star power distribution along with local point-of-load voltage regulators in the new drawer reduces the voltage deviations and the noise coupling along the drawer. Another
important change on the drawer archtecture is the replacement of a single interface board (actual design) by four
Daughter Boards, improving the system robustness and decoupling the drawer electronics into independent units.

\subsubsection{Front-End - Modified 3-in-1 Card}

The Enrico Fermi Institute (University of Chicago) is developing a modified version of the present 3-in-1 card
\cite{shaper}. This front-end Board is composed of discrete components and can be divided in three stages: the fast
signal processing chain, the slow signal processing chain and the calibration electronics and the control bus
interface.

The fast signal processing chain includes a 7-pole passive LC shaper, bi-gain clamping amplifiers with a gain ratio of
32 and a pair of differential drivers feeding the analog signals from the low-gain channel and the high-gain channel to
the ADCs which are placed in the Main Board. The slow signal processing chain includes a programmable 3-gain integrator
which monitors the PMT current induced by a Cesium source and the minimum bias current induced during the collisions.
Finally, the last stage includes a precise charge injection circuit, integrator gain control and the control bus
interface. This modified version has better linearity and a lower noise level than the previous version. The
prototype of the modified 3-in-1 card has been built using COTS components and has passed initial radiation tests. 

The modified 3-in-1 card will equipe the Demonstrator drawer and due to that a different version of the board is currently been produced, including an analog output for the Trigger Board in order to preserve the system compatibility with the current lvl1 triggering system.

\subsubsection{Front-End - QIE Chip}

The Argonne National Laboratory is working on the design of a front-end board which includes a new version of the Charge
(Q) Integrator and Encoder (QIE) chip developed in collaboration with Fermilab and CMS HCAL. The QIE includes
a current splitter composed of 23 splitter transistors, providing 4 different ranges (16/23, 4/23, 2/23, 1/23), followed
by a gated integrator and an on-board 6 bit flash ADC to cover a dynamic range of 17 bits. In this way, only a simple
digital interface is needed to communicate with the Main Board. The QIE also includes a charge injection circuit for
calibration and an integrator for source calibration. The QIE does not perform pulse shaping, minimizing pile up problems and allowing raw PMT pulses to be measured.

At the present time two QIE prototypes have been produced and tested. The first fully functional QIE has already been designed and is under test.

\subsubsection{Front-End - FATALIC ASIC}

FATALIC means Front-End for Atlas TileCal Integrated Circuit \cite{fatalic} which is being designed at Laboratoire de
Physique Corpusculaire in Clermont-Ferrand (LPC). FATALIC includes a multi-gain current conveyor (CC) with three
different gains (1, 8, 64) which cover the full dynamic range of the PMT signal, followed by a shaper in order to
improve the SNR. The readout chain is completed using an external 12 bit pipelined ADC with a
sampling rate of 40~MHz also developed at LPC and called Twelve bits ADC for s-ATLAS TileCal Integrated Circuit
(TACTIC). Moreover FATALIC includes an integrator and a 10 bit ADC with a low sampling rate for calibration purposes.
Both chips are designed using the CMOS IBM 130~nm technology.

The first prototypes of the FATALIC, version 1 and 2, have been produced and tested. The FATALIC version 3, which includes an integrator amplifier, has been
delivered and tests are been performed. The tests with both FATALIC and TATIC are expected by the end of this year.

\subsubsection{Main Board and Daughter Board}

The Main Board \cite{mbdb} is responsible for the digital control of the FEs, data organization and for the
transmission of the data to the Daughter Board. The current prototype design digitizes the signals coming from four
Modified 3-in-1 Cards by using four 12 bit ADCs working at a sampling rate of 40~MHz. The final version of the Main
Board should process 12 PMT signals. 

The Daughter Board \cite{mbdb} is intended to serve as a processing board in the next TileCal electronics drawer and is designed for redundancy and high data throughput readout. The two separately programmable FPGAs are responsible for reading out signals originating from the same tile cells but from different sides of the scintillating tiles. This means that they process equivalent data. If one chain fails it can be replaced by the other (there is a loss of statistics though). The Daughter Boards sends the digitized data to the super Read Out Driver (sROD) via high-speed links using the GBT
protocol \cite{gbt}. In order to perform these functions, the first version of the Daughter Board includes two Xilinx Virtex 6
FPGAs, one $12\times5$~Gb/s transmitter using a SNAP12 format connector and two 5~Gb/s SFP+ format connectors. Future versions of the Daughter Board
will include two Xilinx Kintex 7 FPGAs, one $4\times10$~Gb/s QSFP+ and one transmitter using SNAP12 format connector,  to help selecting the optimum link strategy.

The Daughter and the Main boards have been designed by the Stockholm University and the Enrico Fermi Institute
(University of Chicago). First prototypes of the Main Board and the Daughter Board have been produced and tested. The design of the Main Board for the demonstrator drawer is already finished and the board production started.

\begin{figure}
\begin{center}
\includegraphics[width=7cm]{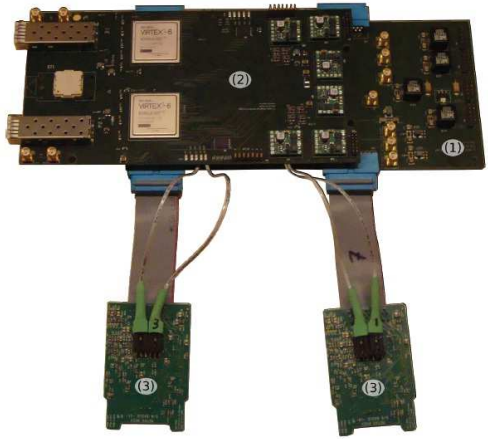}
\end{center}
\caption{\label{prototype}Test mock-up with the new on-detector eletronics prototypes. (1) Main Board, (2) Daughter Board and (3) Front-End Boards (Modified 3-in-1).}
\end{figure}

FIG.~\ref{prototype} shows and early mock-up of the system with only 4 FE inputs. It is possible to see the Main Board prototype (1), which is connected to the Daughter Board prototype (2) and 2 modified 3-in-1 cards (3). 

\subsection{Off-Detector Electronics (sROD)}

\begin{figure*}
\begin{center}
\includegraphics[width=15cm]{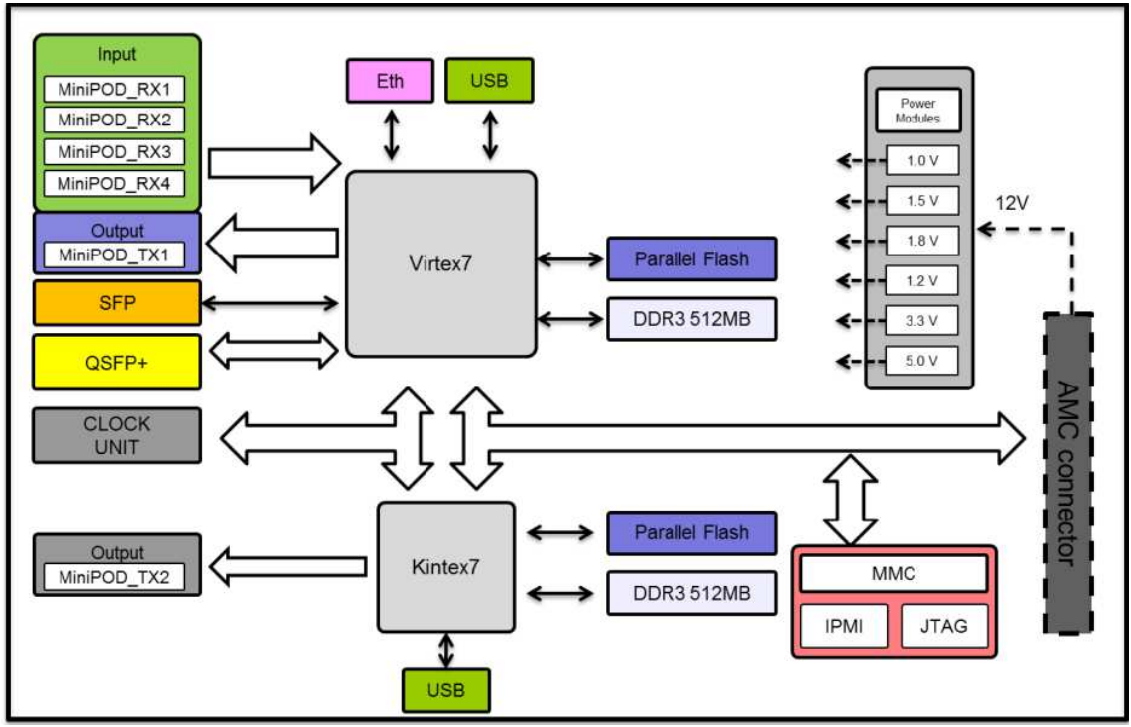}
\end{center}
\caption{\label{srod}sROD Block Diagram.}
\end{figure*}

The sROD demonstrator board \cite{srod} consists of one Xilinx Virtex 7 and one Xilinx Kintex 7 FPGAs as processing
core, 4 receiver Avago MiniPOD connectors, 2 transmitter Avago Mini-POD connectors, one QSFP+ format connector and one
SFP format connector. This board is a double mid-size Advanced Mezzanine Card (AMC) and it is conceived to be plugged in
a Advanced Telecommunications Computing Architecture (ATCA) carrier or in a Micro Telecommunications Computing
Architecture ($\mu$TCA) system. FIG.~\ref{srod} shows a block diagram of the sROD.

The sROD  performs several functions. These include: data processing and data reception from Daughter Boards; Timing,
Trigger and Control (TTC); Detector Control System (DCS) management and transmission to Main Boards; data reconstruction
and transmission to the ReadOut Subsystem (ROS); as well as data preprocessing and transmission to the Level-1 Calorimeter trigger
system (L1Calo).

The Instituto de F\'{\i}sica Corpuscular (IFIC) - CSIC and Universidad de Valencia (UV), the Laborat\'orio de Instrumenta\c{c}\~ao
e F\'{\i}sica Experimental de Part\'{\i}culas (LIP) and the Stockholm University are involved in the development of this
project. The first prototypes of the sROD demonstrator board are expected soon.

\section{Conclusions}
\label{conclusions}

In order to cope with the higher data rates and radiation levels, the on- and off-detector electronics must be
redesigned. For the on-detector eletronics, three different Front-End boards approaches are being considered.
Additionally, in order to provide sufficient data processing, control and interface with the new back-end electronics, a
Main Board and Daughter Board combination are being designed. For the back-end processing, the sROD is under design and
should be ready for tests by the second half of 2013.

A slice of the new upgraded drawer is currently being tested providing promising results. Extensive laboratory tests will be
performed during this year, when one full Demonstrator drawer will be available.

At the end of Phase~0, one Demonstrator drawer should be installed into the detector. The final decision about the
Front-End Board design should be taken by the end of 2018. The production of the new Tilecal readout electronics will
take place during 2019-2020 and should be prepared for installation during Phase~2.

\begin{acknowledgments}
I would like to thank my colleagues from the Tile Calorimeter Group for the fruitful discussions and UFJF, CNPq, CAPES and FAPEMIG from Brazil for the support to this work.
\end{acknowledgments}

%\nocite{*}

%\section*{References}

\bibliography{TileUpgradeLISHEP}% Produces the bibliography via BibTeX.

\end{document}